\documentclass[conference]{IEEEtran}
\IEEEoverridecommandlockouts
\usepackage{cite}
\usepackage{amsmath,amssymb,amsfonts}
\usepackage{algorithmic}
\usepackage{graphicx}
\usepackage{textcomp}
\usepackage{xcolor}
\usepackage{subfigure}
\usepackage{multirow}
\usepackage{makecell}
\usepackage{threeparttable}
\usepackage{colortbl}

\def\BibTeX{{\rm B\kern-.05em{\sc i\kern-.025em b}\kern-.08em
    T\kern-.1667em\lower.7ex\hbox{E}\kern-.125emX}}

\begin{document}
\bstctlcite{IEEEexample:BSTcontrol}

\title{A Homogeneous Processing Fabric for Matrix- Vector Multiplication and Associative Search Using Ferroelectric Time-Domain Compute-in-Memory }

\author{Xunzhao  Yin \IEEEmembership{Member, IEEE}, Qingrong Huang \IEEEmembership{Student Member, IEEE}, Franz Müller \IEEEmembership{Student Member, IEEE}, \\
Shan Deng \IEEEmembership{Student Member, IEEE}, Alptekin Vardar \IEEEmembership{Student Member, IEEE}, Sourav De \IEEEmembership{Member, IEEE}, \\
Zhouhang Jiang \IEEEmembership{Student Member, IEEE}, Mohsen Imani \IEEEmembership{Member, IEEE}, Cheng Zhuo \IEEEmembership{Senior Member, IEEE}, \\Thomas Kämpfe \IEEEmembership{Member, IEEE},
 and Kai Ni \IEEEmembership{Member, IEEE}
\IEEEcompsocitemizethanks{\IEEEcompsocthanksitem X. Yin, Q. Huang, and C. Zhuo are with the College of Information Science and Electronic Engineering, Zhejiang University, Hangzhou, China. E-mail: \{xzyin1, czhuo\}@zju.edu.cn
}
\IEEEcompsocitemizethanks{\IEEEcompsocthanksitem Franz Müller, Alptekin Vardar, Sourav De, and Thomas Kämpfe are with the Fraunhofer IPMS-CNT, Dresden, Germany. E-mail:  thomas.kaempfe@ipms.fraunhofer.de 
}
\IEEEcompsocitemizethanks{\IEEEcompsocthanksitem Shan Deng, Zhouhang Jiang, and Kai Ni are with the Department of Electrical \& Microelectronic Engineering, Rochester Institute of Technology, Rochester, NY. E-mail:{kai.ni@rit.edu}
}

\thanks{}}


\maketitle

\begin{abstract}
In this work, we propose a ferroelectric FET (FeFET) time-domain compute-in-memory (TD-CiM) array as a homogeneous processing fabric for binary multiplication-accumulation (MAC) and content addressable memory (CAM). We demonstrate that: i) the XOR(XNOR)/AND logic function can be realized using a single cell composed of 2FeFETs connected in series; ii) a two-phase computation in an inverter chain with each stage featuring the XOR/AND cell to control the associated capacitor loading and the computation results of binary MAC and CAM are reflected in the chain output signal delay, illustrating full digital compatibility; iii) comprehensive theoretical and experimental validation of the proposed 2FeFET cell and inverter delay chains and their robustness against FeFET variation; iv) the homogeneous processing fabric is applied in hyperdimensional computing to show dynamic and fine-grain resource allocation to accommodate different tasks requiring varying demands over the binary MAC and CAM resources.
\end{abstract}

\vspace{-2ex}
\section{Introduction}
\label{sec:intro}
\begin{figure}[t]
\centering
{\includegraphics[width=0.95\columnwidth]{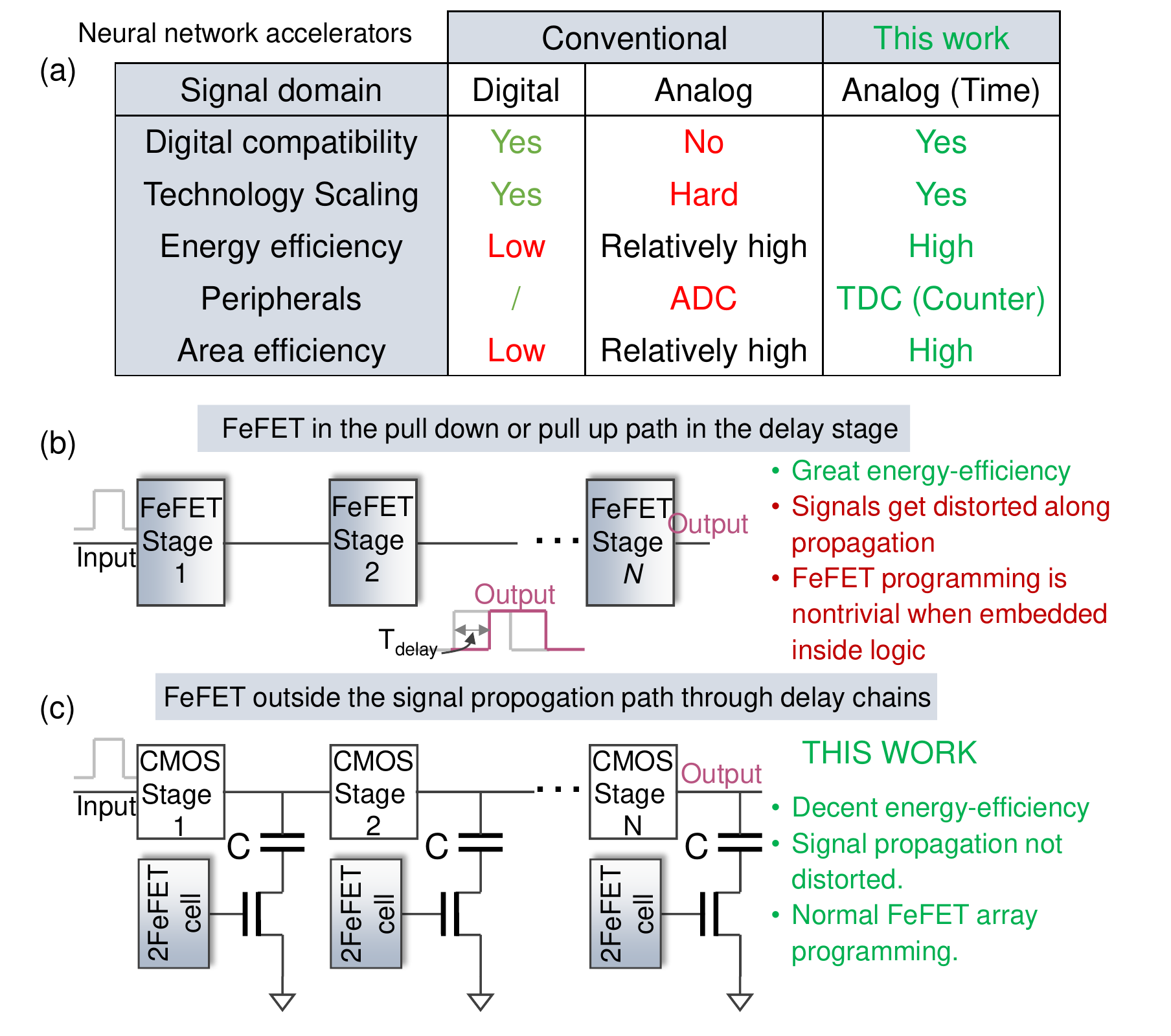}}
\vspace{-2ex}
\caption{(a) Time domain compute-in-memory is promising for its simplicity. (b) TD-CiM using FeFETs in the signal transmission path achieves high energy-efficiency while remains several issues. (c) This work proposes TD-CiM with FeFET outside the signal transmission path.  }
\vspace{-4ex}
\label{fig:digital_analog_CiM}
\end{figure}

For the hardware acceleration of neural networks to support wide deployment of intelligent devices, different approaches have been proposed and studied. 
The commercialized digital processors and near-memory computing processors are  have gained tremendous performance boost over the years through technology scaling \cite{keshavarzi2020ferroelectronics}. The main challenge, however, for these computing platforms is the power-hungry and slow memory access, as summarized in Fig.\ref{fig:digital_analog_CiM}(a). 
Analog approaches, such as CiM, are proposed to address this issue by directly performing the computations inside the memory array, thus eliminating the data transfer bottleneck \cite{keshavarzi2020ferroelectronics, sebastian2020memory, ielmini2018memory, ielmini2020device, zhang2020neuro}. As the examples, the crossbar array is used to  perform the MAC operation \cite{sebastian2020memory} to accelerate matrix-vector multiplication, and CAM designs have been used for parallel and efficient associative search \cite{ni2019ferroelectric, hu2021memory, yin2022ferroelectric, yin2022ultra, kazemi2021fefet}. 
Though these designs are promising, they still suffer from typical digital incompatibility and costly peripherals to convert signals between digital and analog domains.

TD-CiM is emerging as an alternative computing paradigm to address the challenges associated with the pure digital and analog processors \cite{yang2021timaq}. Its core is a digital delay chain, whose delay is controlled in an analog fashion using memory cells. The computation results are reflected in the signal propagation delay, sensing which only needs digital techniques. 
TD-CiM is compact, energy-efficient, and compatible with digital circuits and associates with simple peripheral sensing circuitry. Recently, by leveraging compact, energy-efficient, and CMOS compatible HfO\textsubscript{2} based FeFETs, TD-CiM array is proposed \cite{luo2021energy}, which is an inverter chain composed of \textit{N} stages each embedding a FeFET device in the pull-down path, as shown in Fig.\ref{fig:digital_analog_CiM}(b). Such a design achieves remarkable energy-efficiency, while also has many unaddressed issues. 
Since the FeFETs participate in the signal propagation path as a tunable resistance, the final output delay is highly sensitive to the FeFET variation. In addition, the large ON/OFF ratio of FeFET can not be fully exploited for the output delay modulation as a FeFET exhibiting an OFF state may prevent the signal propagation, resulting in a computation failure. Therefore, the conductance range of FeFET that can be used in the design is limited. Moreover, with the FeFETs inside the signal transmission path, it is challenging to write FeFETs, especially in the array settings where inhibition bias schemes need to be applied.
To address these issue, we propose a novel FeFET based TD-CiM where FeFETs are outside the signal propagation path, but rather controls the associated capacitor loading, as shown in Fig.\ref{fig:digital_analog_CiM}(c), thus regaining full digital compatibility and ease to operate.

\begin{figure}[h]
\centering
\vspace{-2ex}
{\includegraphics[width=0.9\columnwidth]{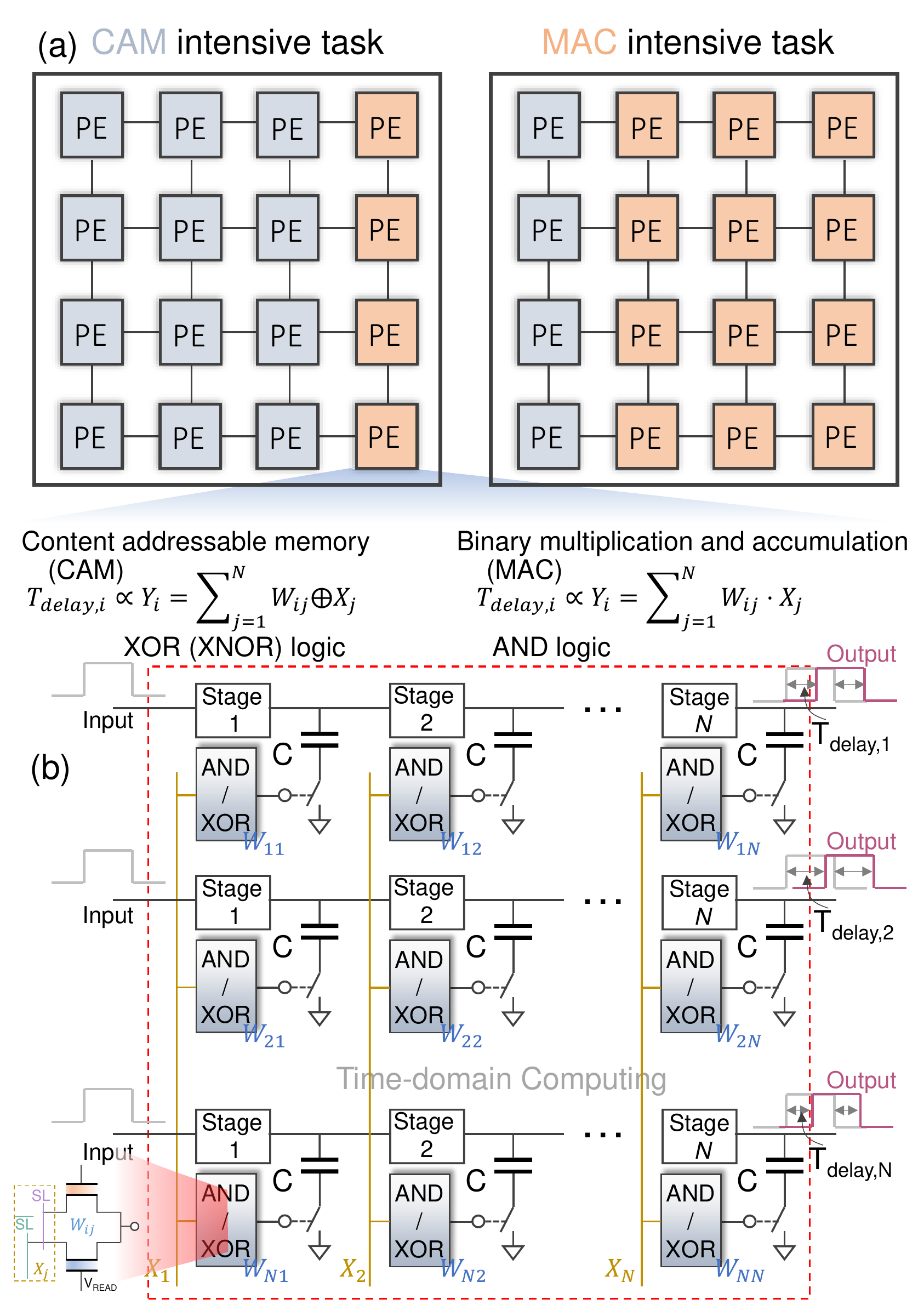}}
\vspace{-2ex}
\caption{(a) A homogeneous PE array that can function as both binary CAM (in gray) and MAC (in orange). It can be dynamically allocated to either CAM or MAC to adjust to various task workloads. (b) The proposed FeFET based TD-CiM to realize both binary CAM and MAC. }
\vspace{-2ex}
\label{fig:homogeneous_array}
\end{figure}

In addition, our proposed FeFET based TD-CiM enables a homogeneous processing fabric featuring that it can serve as either binary MAC array through Boolean AND logic operation or CAM through XOR/XNOR logic operation using the same cell and array structure. MAC and CAM are important CiM design macros, and combining them together suggests wide data-intensive applications, such as the memory augmented neural network \cite{ni2019ferroelectric} or hyperdimensional computing (HDC) \cite{imani2019binary}. 
With the proposed fabric, it is expected that the same chip can be configured to process different tasks/workloads that require varying demands on the amounts of the MAC and CAM through dynamic and fine-grain resource allocation, as illustrated in Fig.\ref{fig:homogeneous_array}(a). 
The fabric is realized through inverter chains with a load capacitor at each stage, as shown in Fig.\ref{fig:homogeneous_array}(b). The load capacitor is used to involve in the signal propagation, and controlled through an AND/XOR cell, which can be realized with a 2FeFET cell. 
In our proposed design, the FeFETs are outside the signal propagation path, and the 2FeFET AND/XOR cell stores binary values by leveraging the non-volatile memory property of FeFET, therefore, the design is robust against device variation, and maintains
the signal propagation integrity. 
In addition, with access to the FeFETs source/drain, it is possible to program the FeFETs with write inhibition schemes \cite{ni2018write, xiao2022on}. 
In the following, the 2FeFET cell implementing both the AND and XOR(XNOR) logic operation is introduced and validated in section \ref{sec:cell}. Then the FeFET TD-CiM array using the 2FeFET cell is built and validated in section \ref{sec:CiM_design} and section \ref{sec:evaluation}, respectively. In section \ref{sec:app_benchmarking}, the proposed FeFET TD-CiM is applied for hyperdimensional computing. Section \ref{sec:conclusion} concludes the paper.    

\vspace{-1ex}
\section{2FeFET Cell for XOR \& AND Operation}
\label{sec:cell}

\begin{figure*}[t]
	\centering
	\vspace{-1ex}
	\includegraphics[width=1\linewidth]{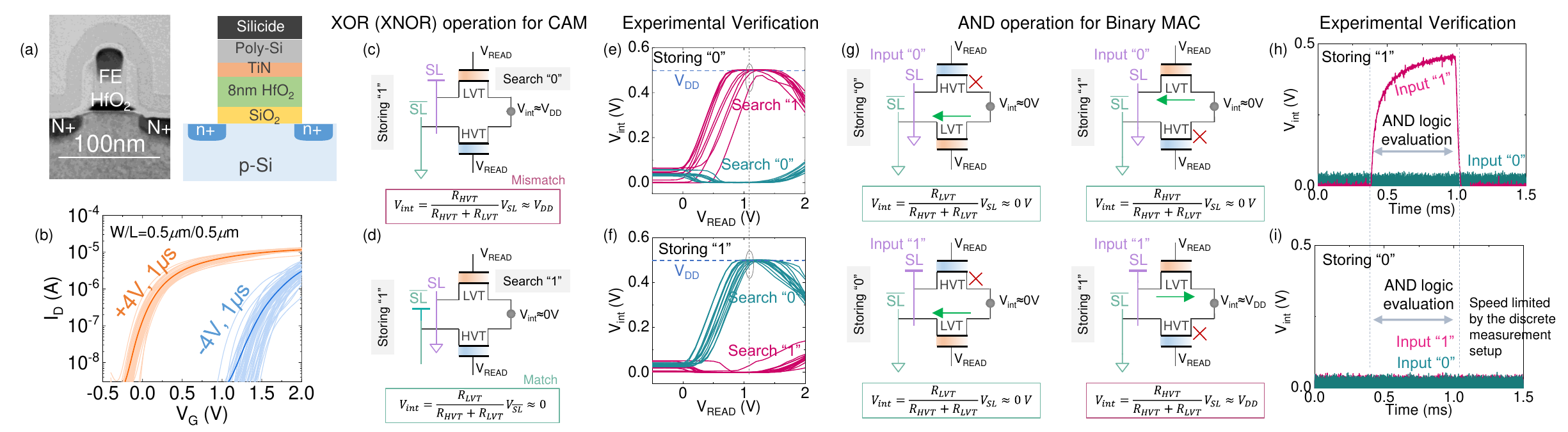}
	\vspace{-2ex}
	\caption{(a) TEM and schematic cross section of tested FeFETs integrated on 28nm HKMG platform. (b) \textit{I}\textsubscript{D}-\textit{V}\textsubscript{G} characteristics for the low-\textit{V}\textsubscript{TH} and the high-\textit{V}\textsubscript{TH} states for 50 different devices. XOR (XNOR) logic implementation using the 2FeFET cell for (c) mismatch and (d) match scenarios, respectively. Experiments verify the XOR(XNOR) logic when storing (e) "0" and (f) "1", respectively. 17 cells are tested. (g) Illustration of the AND operation in a 2FeFET cell under different input operand and stored operand configurations. Experiments verify the AND operation when storing (h) "1" and (i) "0", respectively.    }
	\label{fig:fig3_2fefetcell}
	\vspace{-2ex}
\end{figure*}

In this work, FeFETs integrated on industrial 28nm high-$\kappa$ metal gate platform as shown in Fig.\ref{fig:fig3_2fefetcell}(a) are demonstrated and used. The device features a gate stack of poly-silicon/TiN/doped HfO\textsubscript{2}/SiO\textsubscript{2}/p-Si, and detailed process information can be found in \cite{trentzsch201628nm}. 
When subjected to $\pm$4V, 1$\mu$s write pulse, the FeFET will be set to the low-\textit{V}\textsubscript{TH} state or the high-\textit{V}\textsubscript{TH} state, respectively, as shown in Fig.\ref{fig:fig3_2fefetcell}(b). For the tested 50 devices with a size W/L=0.5$\mu$m/0.5$\mu$m, a large memory window between the two states can still be obtained, indicating reasonable device variation control of FeFET \cite{beyer2020fefet}.

Fig.\ref{fig:fig3_2fefetcell}(c) and (d) show the operation principles of the 2FeFET cell structure to implement the XOR(XNOR) logic which is the key function in a CAM cell for match and mismatch conditions, respectively. The 2FeFETs are connected in series, forming a push-pull structure. By writing complementary states, i.e., low-\textit{V}\textsubscript{TH} state and high-\textit{V}\textsubscript{TH} state, into FeFETs to encode the stored information and then applying the search information on the search line (SL and $\overline{\text{SL}}$), a match or mismatch between the stored information and applied search input can be determined. For example, for a cell storing bit "1", as shown in Fig.\ref{fig:fig3_2fefetcell}(c), the upper and lower FeFETs are written to the low-\textit{V}\textsubscript{TH} state and high-\textit{V}\textsubscript{TH} state, respectively. Then a bit “0” is searched by applying \textit{V}\textsubscript{DD}/GND to SL/$\overline{\text{SL}}$. In this way, due to the resistor divider formed by the two FeFETs, the internal node voltage, \textit{V}\textsubscript{int}, is given as

\begin{equation}
    V_{int}=\frac{R_{HVT}}{R_{HVT}+R_{LVT}}V_{SL}
\end{equation}

\noindent where the \textit{R}\textsubscript{HVT}/\textit{R}\textsubscript{LVT} is the equivalent FeFET channel resistance for the high/low-\textit{V}\textsubscript{TH} state. The internal node will be charged to a high voltage through the upper  FeFET with low-\textit{V}\textsubscript{TH}, indicating a mismatch. On the other hand, flipping the search information to bit "1" corresponds to applying GND/\textit{V}\textsubscript{DD} to SL/$\overline{\text{SL}}$, as shown in Fig.\ref{fig:fig3_2fefetcell}(d). In this case, the \textit{V}\textsubscript{int} is given by 

\begin{equation}
    V_{int}=\frac{R_{LVT}}{R_{HVT}+R_{LVT}}V_{\overline{SL}}
\end{equation}

The internal node will be discharged to ground through the upper  FeFET with low-\textit{V}\textsubscript{TH}, indicating a match. Similar analysis can be performed when the stored information changes to bit "0", where the upper and lower FeFETs are written into the high-\textit{V}\textsubscript{TH} state and low-\textit{V}\textsubscript{TH} state, respectively. Based on this 2FeFET cell, XOR(XNOR) logic can be realized.

To verify the cell operation, 2FeFET cell is constructed experimentally and characterized. Fig.\ref{fig:fig3_2fefetcell}(e) and (f) show the \textit{V}\textsubscript{int} as a function of applied read gate voltage, \textit{V}\textsubscript{READ}, during search when the cell stores bit "0" and "1", respectively. When the cell stores bit "0" and a search input "1" is applied, as shown in Fig.\ref{fig:fig3_2fefetcell}(e), the \textit{V}\textsubscript{int} grows initially with the \textit{V}\textsubscript{READ} and starts to decrease with continual increase of \textit{V}\textsubscript{READ}. This is because that to read out the the value stored in the FeFET, an appropriate \textit{V}\textsubscript{READ} between the low-\textit{V}\textsubscript{TH} and high-\textit{V}\textsubscript{TH} is needed. When \textit{V}\textsubscript{READ} is high enough such that even the FeFET storing the high-\textit{V}\textsubscript{TH} state is turned ON, the \textit{V}\textsubscript{int} starts to decrease with the \textit{V}\textsubscript{READ}. With an appropriate \textit{V}\textsubscript{READ}, e.g., 1V, the \textit{V}\textsubscript{int}s for the match and mismatch scenarios are widely separated, even considering the device variations among 17 measured cells. The robustness of the 2FeFET cell against device variation is another advantage, which will be further studied in section \ref{sec:evaluation}.

Using the same 2FeFET cell, the AND logic function can also be realized. In this case, the encoding of the stored information, which is one of the operands, remains the same as the XOR(XNOR) case by storing complementary \textit{V}\textsubscript{TH} states into the two FeFETs. The other operand, i.e., the input voltages on the SL/$\overline{\text{SL}}$, however, are different from the XOR(XNOR) case. The input voltage is only applied on the SL while the $\overline{\text{SL}}$ is fixed at GND. 
Fig.\ref{fig:fig3_2fefetcell}(g) shows all four scenarios corresponding to different combinations of the two operands. When bit "0" is stored, the lower FeFET is with the low-\textit{V}\textsubscript{TH} state, and will discharge the internal node upon the operation, irrespective of the input SL voltage. 
When the cell stores bit "1", an input "0" causes the internal node voltage to discharge through the upper  FeFET with the low-\textit{V}\textsubscript{TH} state. The only scenario that will yield a high \textit{V}\textsubscript{int} is when the cell stores "1" and input "1" is applied such that the \textit{V}\textsubscript{int} is charged up through the upper FeFET with the  low-\textit{V}\textsubscript{TH} state. In this way, the AND logic is realized. 
The cell operations for the AND logic are also experimentally verified. Fig.\ref{fig:fig3_2fefetcell}(h) and (i) show the transient waveform of \textit{V}\textsubscript{int} for cell storing “1” and “0”, respectively. The results show that the \textit{V}\textsubscript{int} is only high when the AND logic output is true and low otherwise, validating the AND logic operation. Note that the delay shown in the waveform is limited by the discrete measurement setup that is adopted and fully integrated cell is expected to operate much faster, which will be studied using SPICE simulation in section \ref{sec:CiM_design}.

\vspace{-1ex}

\section{Principles of FeFET Time-Domain CiM}
\label{sec:CiM_design}

\begin{figure}[h]
\centering
\vspace{-2ex}
{\includegraphics[width=0.95\columnwidth]{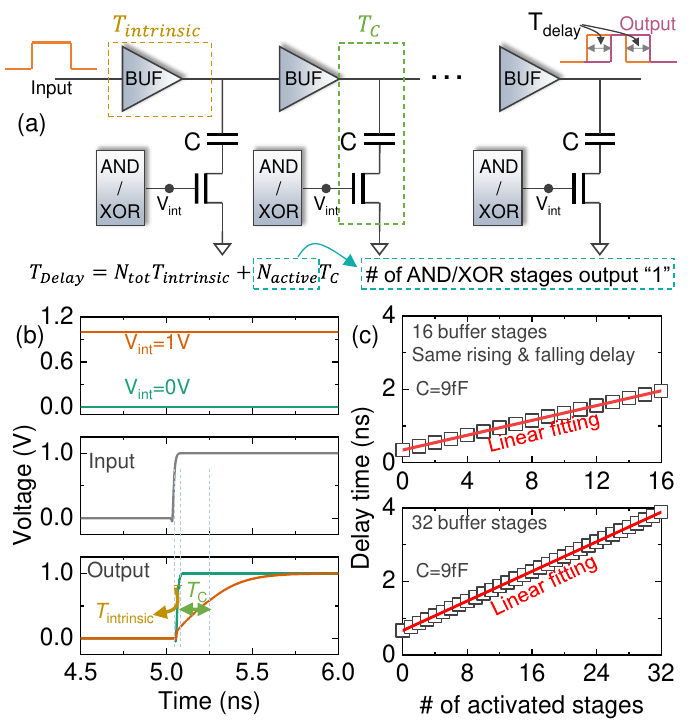}}
\vspace{-2ex}
\caption{(a) Total delay of an input pulse through a buffer chain is determined by the number of  AND/XOR cells with '1' outputs, which activate the loading capacitor of the corresponding stage. (b) Transient waveforms of a single buffer stage. (c) The delay is linearly dependent on the number of activated stages, thus realizing AND/XOR operation in time domain.}
\vspace{-2ex}
\label{fig:buffer_chain}
\end{figure}

By embedding the 2FeFET cell in section \ref{sec:cell} into the array shown in Fig.\ref{fig:digital_analog_CiM}(c), the proposed FeFET based TD-CiM is realized.  To understand the operation principles of the proposed FeFET TD-CiM, the buffer delay chains are first evaluated, as shown in Fig.\ref{fig:buffer_chain}(a), which features the symmetric rising edge and falling edge delay. An input pulse is sent to the first stage and the corresponding output pulse thus has a delay, \textit{T}\textsubscript{delay}, with respect to the input pulse. 
In addition to the intrinsic delay of each buffer, \textit{T}\textsubscript{intrinsic}, \textit{T}\textsubscript{delay} also contains the delay caused by the conditionally loaded capacitors for each stage. 
The capacitor loading occurs when the AND/XOR(XNOR) logic of the corresponding stage outputs a high \textit{V}\textsubscript{int}, which turns on the access transistor and activates the load capacitor. In this sense, \textit{T}\textsubscript{delay} can be expressed as:

\begin{equation} \label{eq:delay}
    T_{delay} = N_{tot}T_{intrinsic}+N_{active}T_{C}
\end{equation}

\noindent where the \textit{N}\textsubscript{tot} and \textit{N}\textsubscript{active} are the number of total stages and the active stages where the AND/XOR logic outputs "1", respectively. \textit{T}\textsubscript{C} is the additional delay incurred by the capacitor loading. Sensing the \textit{T}\textsubscript{delay} allows to detect \textit{N}\textsubscript{active}, which reflects the outputs of the XOR(XNOR) and AND logic with in the stages, thus realizing the binary MAC and CAM operations, respectively.

\begin{figure}[h]
\centering
\vspace{-2ex}
{\includegraphics[width=1\columnwidth]{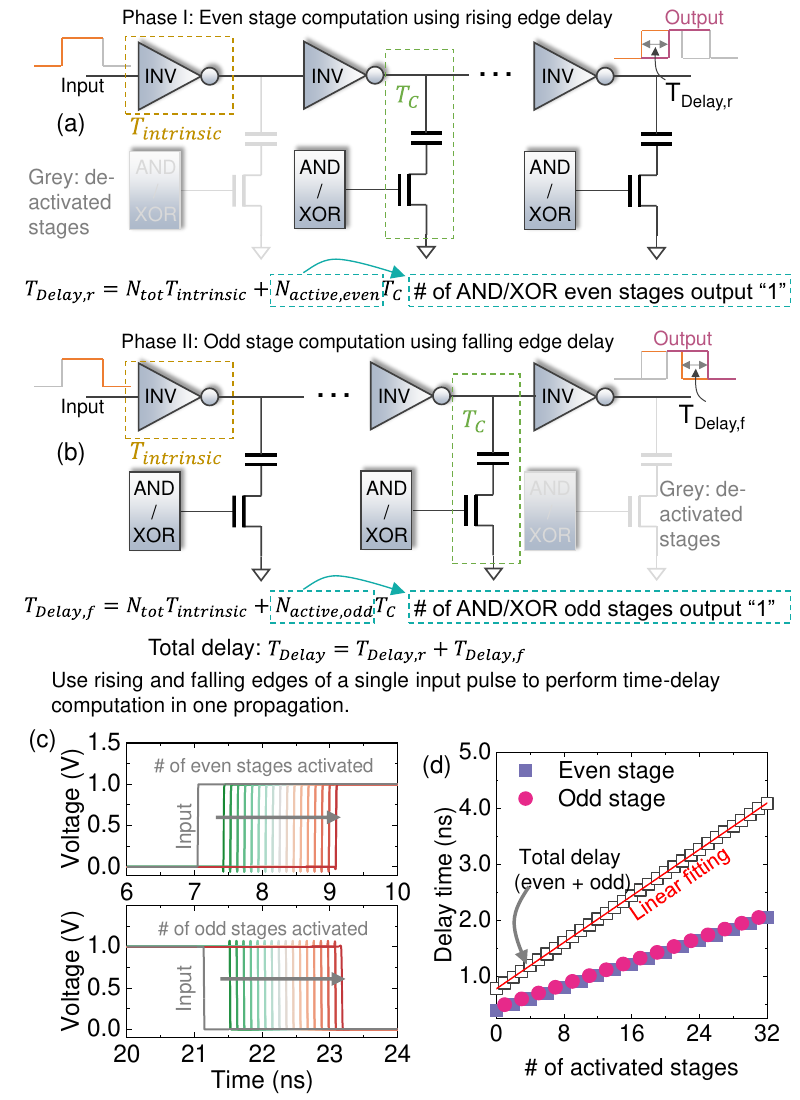}}
\vspace{-4ex}
\caption{Using inverter chains avoids the density penalty in the buffer chain. Computation is completed in 2 phases, where only (a)/(b) even/odd stages are activated in phase I/II, and the delay results are measured from the rising/falling edges. (c) Output transients with varying number of activated even or odd stages. (d) Evaluated delay shows a linear dependence on the number of activated stages. 
For the even and odd stage delay, the x-axis represents the index of the activated stage within the chain, and the stages before the index are activated. 
}
\vspace{-2ex}
\label{fig:inverter_chain}
\end{figure}

To verify the working principles, a compact multi-domain FeFET model~\cite{ni2018circuit} and a 40nm MOSFET Predictive Technology Model (PTM) \cite{ASUmodel} with minimized transistor sizes have been adopted in the SPICE simulation. The load capacitor of each stage is set to 9fF unless explicitly mentioned.
Fig.\ref{fig:buffer_chain}(b) shows the transient waveform of a single buffer stage. 
When the \textit{V}\textsubscript{int}=0V, the associated load capacitor is deactivated and the delay between the output and input pulses is \textit{T}\textsubscript{intrinsic}. 
When the load capacitor is activated, the additional delay, \textit{T}\textsubscript{C}, caused by the load capacitor contributes to the total delay. 
Fig.\ref{fig:buffer_chain}(c) shows the \textit{T}\textsubscript{delay} of 16 and 32 buffer stages with varying number of activated stages to evaluate the capability of a buffer chain in realizing the binary MAC/CAM operation. A linear dependence of \textit{T}\textsubscript{delay} on the \textit{N}\textsubscript{active} is validated, which is consistent with  Eq.\eqref{eq:delay}. Therefore, sensing the \textit{T}\textsubscript{delay} allows to back calculate the \textit{N}\textsubscript{active}, thus realizing the binary MAC/CAM operation.

Building on the buffer delay chains, a denser solution using inverter chains is proposed, as shown in Fig.\ref{fig:inverter_chain}. Without degrading the signal shapes during the signal propagation, a two-phase operation is proposed. 
In the phase I, as shown in Fig.\ref{fig:inverter_chain}(a), all the odd stages AND/XOR cell output zero, thus disabling the odd stages capacitors. Then only the load capacitors of even stage participate in the computation, making the chain effectively a buffer chain. The rising edge delay of the output pulse compared with the input pulse is given by

\begin{equation} \label{eq:delayr}
    T_{delay,r} = N_{tot}T_{intrinsic}+N_{active, even}T_{C}
\end{equation}

\noindent where \textit{N}\textsubscript{active, even} is the number of activated even stages. Similarly for the phase II, as shown in Fig.\ref{fig:inverter_chain}(b), all the even stages are disabled and the total delay includes the delay contributed by the activated odd stages. To save energy, it is possible to apply a single input pulse propagation to perform the computation, where the rising and fall edges are sensed for computation in phase I and II, respectively. This requires the input pulse width to be larger than the phase I delay such that the signal propagation of phase I completes before the activated stages switch from even to odd, otherwise the phase I delay in Eq. \eqref{eq:delayr} is not valid.
In this way, the output falling edge delay with respect to the input pulse is given by

\begin{equation} \label{eq:delayf}
    T_{delay,f} = N_{tot}T_{intrinsic}+N_{active, odd}T_{C}
\end{equation}

\noindent where \textit{N}\textsubscript{active, odd} is the number of activated odd stages. Then the total delay \textit{T}\textsubscript{delay}, i.e., the results for MAC/CAM function is obtained by adding the two phase delay results together.
\begin{equation} \label{eq:delayt}
    T_{delay} = T_{delay,r}+T_{delay,f}
\end{equation}

Fig.\ref{fig:inverter_chain}(c) shows the simulated rising and falling edge waveforms for phase I and II operations in an inverter chain with 32 stages. The evaluated delays for even and odd stages are shown in Fig.\ref{fig:inverter_chain}(d). The linear delay increase with the number of activated stages and the total delay shows a similar dependence as the buffer chain shown in Fig.\ref{fig:buffer_chain}(c).

\vspace{-1ex}
\begin{figure}[h]
\centering
{\includegraphics[width=1\columnwidth]{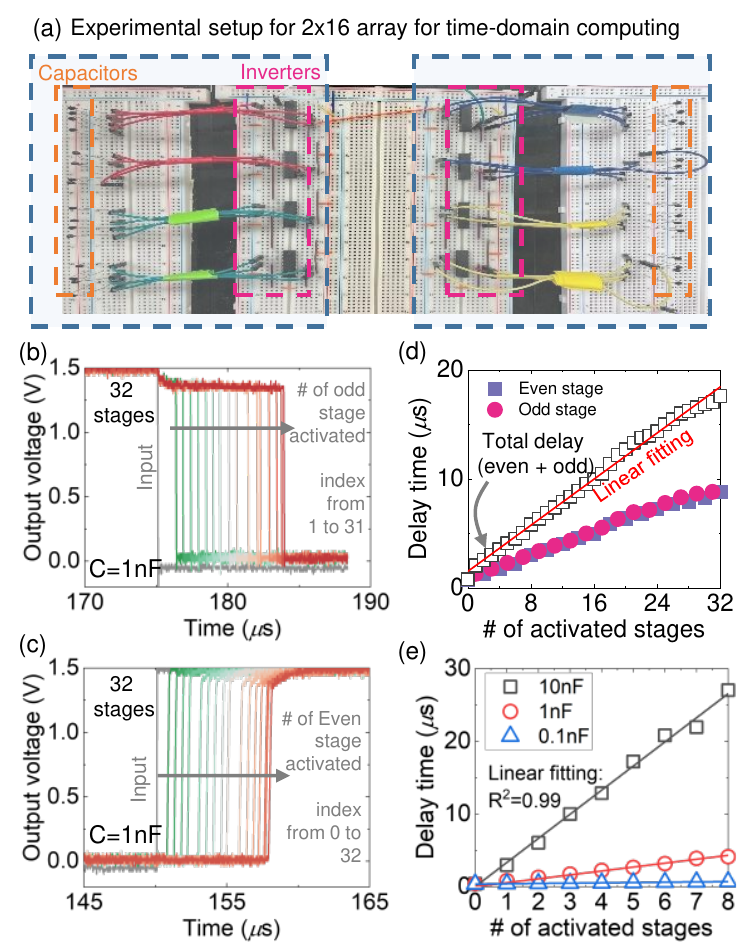}}
\vspace{-4ex}
\caption{Experimental verification of the inverter chain based TD-CiM. (a) Experimental setup (e.g., 2x16 array). (b)/(c) Output transient waveforms for odd/even stages computation in a 32 stages inverter chain. (d) A linear dependence of the delay on the number of activated stages is observed, validating the TD-CiM operation. (e) Dependence of the delay time on the load capacitor size. The linear relationship is maintained for different capacitor sizes.}
\vspace{-4ex}
\label{fig:tdcim_exp}
\end{figure}
\section{Verification of FeFET TD-CiM}
\label{sec:evaluation}
Fig.\ref{fig:tdcim_exp}(a) shows the setup of a 2x16 inverter arrays using discrete inverters and load capacitors to experimentally validate the proposed FeFET TD-CiM.
Fig.\ref{fig:tdcim_exp}(b) and (c) show the transient waveforms of the falling and rising edges  in an inverter chain with 32 stages. The falling edge is sensed to perform the  computation associated with the odd stages while the rising edge is sensed to perform the  computation associated with the even stages. Similar to the simulation results shown in Fig.\ref{fig:inverter_chain}(c), the delay of the falling or the rising edge increases with the number of activated odd or even stages. Evaluated delay results shown in Fig.\ref{fig:tdcim_exp}(d) suggest the linear dependence of the total delay on the number of activated stages. In addition, different load capacitor sizes have been tested for an inverter chain with 8 stages, as shown in Fig.\ref{fig:tdcim_exp}(e). The linear dependence of the total delay on the number of activated stages also holds for different load capacitor values.
Note that these experimental results are intended for functionality verification. The speed and the load capacitors, due to the discrete measurement setup, are limited. It is expected that for fully integrated implementations, a much smaller capacitor can be used to improve the energy-efficiency and speed.

\begin{figure*}[t]
	\centering
	\vspace{-1ex}
	\includegraphics[width=1\linewidth]{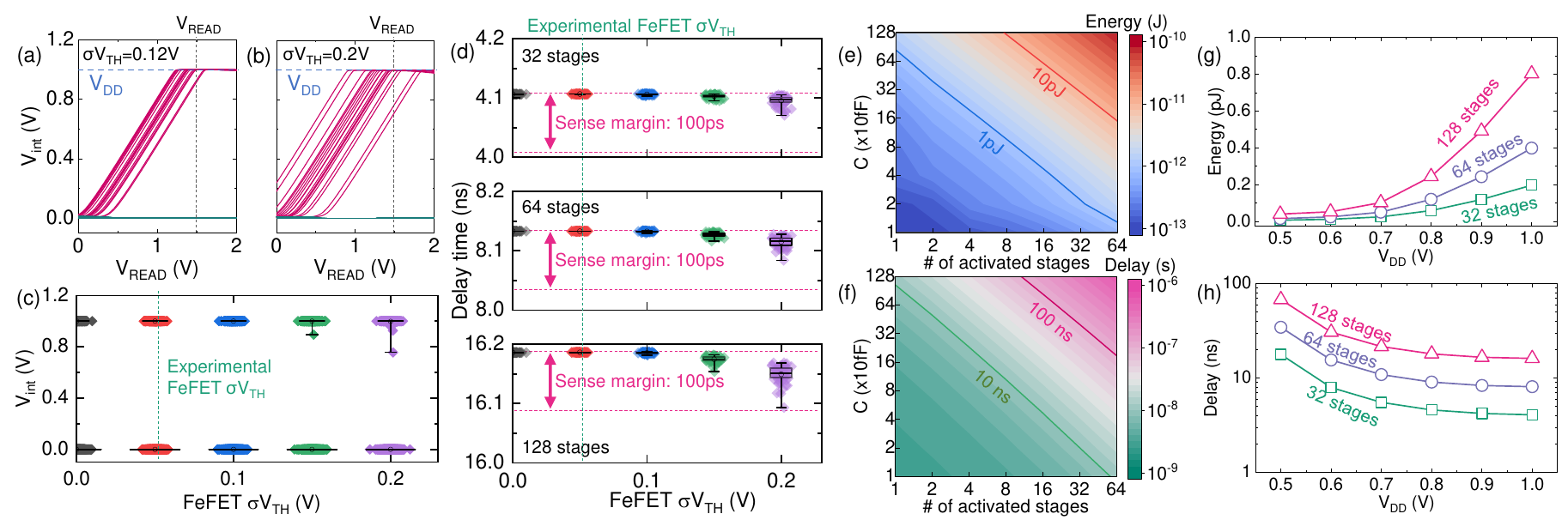}
	\vspace{-4ex}
	\caption{Internal node voltage variation versus \textit{V}\textsubscript{READ} with the FeFET \textit{V}\textsubscript{TH} variation of (a) $\sigma$\textit{V}\textsubscript{TH}=0.12V and (b) $\sigma$\textit{V}\textsubscript{TH}=0.2V. (c) \textit{V}\textsubscript{int} distribution with different $\sigma$\textit{V}\textsubscript{TH} values. (d) Sensing the computation results, i.e., the delay time exhibit sufficient sense margins due to the series structure of 2FeFET cell. (e) Energy and (f) delay of the proposed TD-CiM array with varying load capacitance in varying array sizes. (g) Energy and (h) delay of the arrays with different sizes under supply voltage scaling.}
	\label{fig:variation_dse}
	\vspace{-3ex}
\end{figure*}

As mentioned in section \ref{sec:cell}, the 2FeFET cell structure exhibits superior resilience against FeFET variations. This is because that as long as the resistance ratio between the high-\textit{V}\textsubscript{TH} state and low-\textit{V}\textsubscript{TH} state is large enough (e.g., $>$10), the internal node voltage, \textit{V}\textsubscript{int}, will reach closely to \textit{V}\textsubscript{DD} or GND depending on the logic output. 
Here we further study the impact of the FeFET \textit{V}\textsubscript{TH} variation on the delay chain operation through Monte Carlo simulations.
We first verify the robustness of the 2FeFET cell by measuring the internal node voltage \textit{V}\textsubscript{int} by 60 Monte Carlo runs assuming a \textit{V}\textsubscript{TH} variation $\sigma$\textit{V}\textsubscript{TH}.
Fig.\ref{fig:variation_dse}(a) and (b) shows the \textit{V}\textsubscript{int} as a function of the read gate voltage, \textit{V}\textsubscript{READ}, for the FeFETs with $\sigma$\textit{V}\textsubscript{TH}=0.12V and $\sigma$\textit{V}\textsubscript{TH}=0.2V, respectively, which is similar to the experimental results shown in Fig.\ref{fig:fig3_2fefetcell}(e) and (f). 
As shown, larger $\sigma$\textit{V}\textsubscript{TH} enlarges the distribution of \textit{V}\textsubscript{int}. Fig.\ref{fig:variation_dse}(c) summarizes the \textit{V}\textsubscript{int} distributions upon \textit{V}\textsubscript{READ} under different $\sigma$\textit{V}\textsubscript{TH} values. 
Note that  a fabricated FeFET with a size of W/L=500nm/500nm is experimentally measured to exhibit around 50 mV $\sigma$\textit{V}\textsubscript{TH} \cite{soliman2020ultra}, which is far less than the assumed $\sigma$\textit{V}\textsubscript{TH}. 
Therefore,  the 2FeFET cell output has superior robustness against the device variation.    
The impact of cell level variation on the accuracy of inverter chain operation is studied in Fig.\ref{fig:variation_dse}(d), where 32, 64, and 128 stages are considered. 
When the FeFET $\sigma$\textit{V}\textsubscript{TH} or the number of stages increases, the distribution of the inverter delay time becomes wider.
Considering a sense margin of 100ps, the proposed FeFET TD-CiM is highly robust to the FeFET variation.

The energy and latency metrics of the proposed FeFET TD-CiM with respect to different load capacitor values varying from 10fF to 1280fF and different number of stages varying 1 to 64 are illustrated in Fig.\ref{fig:variation_dse}(e) and (f). 
The contour lines corresponding to a fixed energy consumption or latency are in the diagonal direction, indicating that the energy and delay are both proportional to the product of the load capacitor value and the number of activated stages (i.e., the total capacitance participating in the computation).
Therefore a small load capacitor is preferred. 
However, the sensing circuitry, such as a counter, which is required to distinguish the small delay contributed by the small load capacitor, \textit{T}\textsubscript{C}, limits the lower bound of the load capacitor value. 
Moreover, the impact of the supply voltage, \textit{V}\textsubscript{DD}, on the array energy and delay metrics is investigated and shown in Fig.\ref{fig:variation_dse}(g) and (h), respectively. 
It can be seen that the proposed array energy decreases and delay increases as the supply voltage scales down.

\vspace{-1ex}
\section{Application of FeFET TD-CiM for Hyperdimensional Computing}
\label{sec:app_benchmarking}

\begin{figure}[h]
\centering
{\includegraphics[width=0.9\columnwidth]{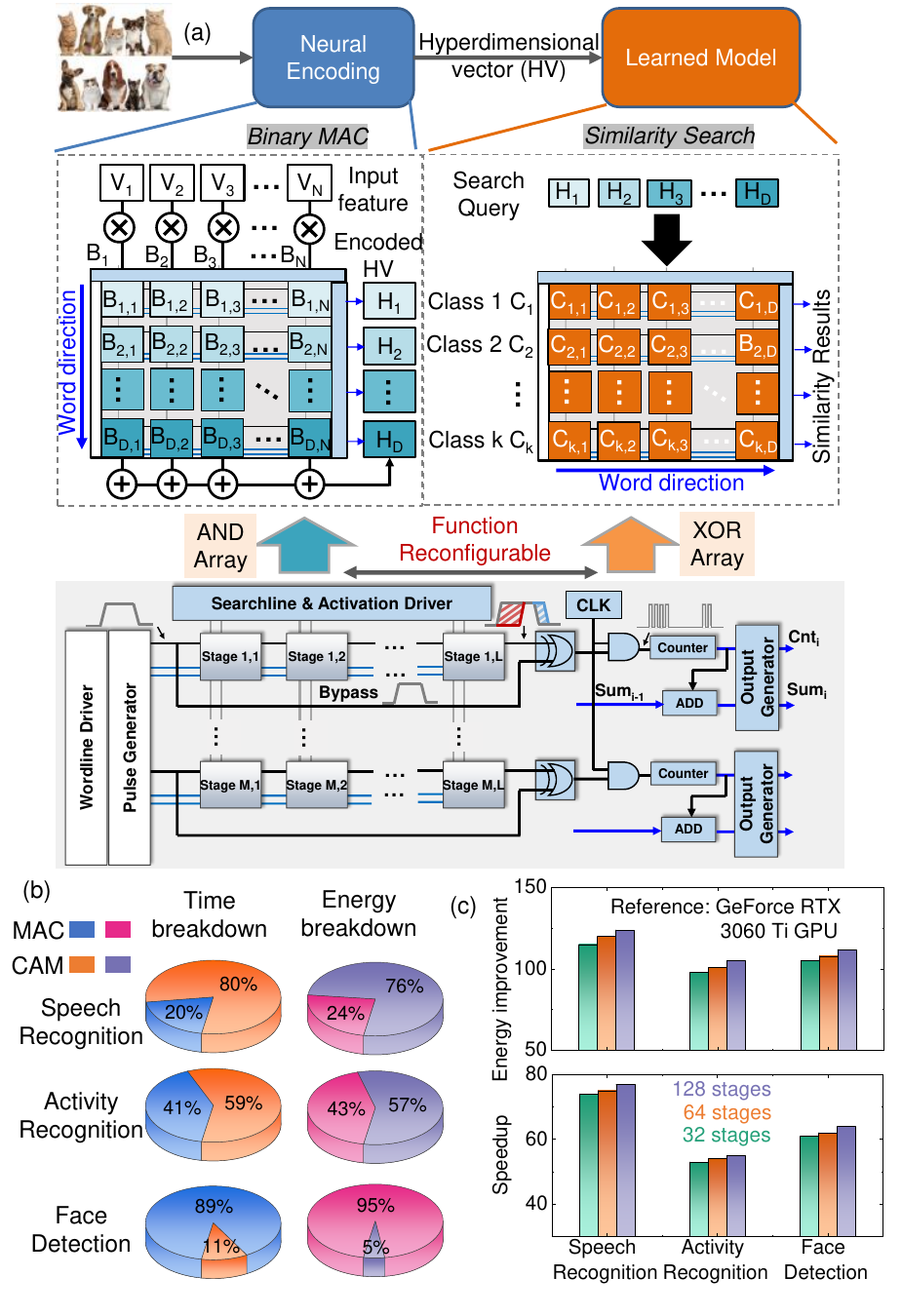}}
\vspace{-2ex}
\caption{(a) The HDC system implementation by configuring the FeFET based TD-CiM to binary MAC and CAM functions depending on the HDC encoding, learning and inference. (b) The percentage of resource required for different information process tasks. (c) Energy improvement and speedup of our FeFET based TD-CiM compared with the standard GPU. }
\vspace{-2ex}
\label{fig:hdc}
\end{figure}

As illustrated in Fig.\ref{fig:homogeneous_array}, our proposed FeFET TD-CiM processing fabric can perform both binary MAC and CAM operations, thus allowing to adapt to different information processing tasks demanding different workloads of MAC and CAM. 
As a case study, we benchmark our design in the hyperdimensional computing (HDC) paradigm, as shown in Fig.\ref{fig:hdc}. 
HDC emerges as an alternative paradigm that mimics the critical brain dynamics for high-efficiency and noise-tolerant computation \cite{imani2019searchd}. 
It is motivated by the observation that the human brain operates on high-dimensional data representations, and is robust against noise. 
As shown in Fig.\ref{fig:hdc}(a), the HDC inference is composed of encoding module and similarity search module.
This encoding method, inspired by the Radial Basis Function (RBF) kernel trick~\cite{rahimi2007random}, considers the non-linear relation between the features during the encoding, and maps the input feature data points into high-dimensional space.
Considering an encoding function that maps a feature vector $\vec{F}=\{V_1,~V_2, \dots,~V_N\}$, with $N$ features  to a hypervector $\vec{H}=\{H_1,~H_2, \dots,~H_D\}$ with $D$ dimensions ($H_i \in \{0,1\}$).
Each dimension of the encoded data is generated by calculating the matrix-vector multiplication of a feature vector with a base matrix $\{\vec{B}_1, \vec{B}_2, \cdots, \vec{B}_N\}$, i.e., $\vec{H}= \sum_{i \in N} V_i \times \vec{B}_i$, where $B_i$ is a randomly generated vector from a set $\{0, 1\}$ with the same dimensionality of the feature vector $\vec{H}$. 
The random vectors $\{\vec{B}_1, \vec{B}_2, \cdots, \vec{B}_N\}$ can be generated once offline and then used for the rest of the inference task ($\vec{B}_i \in \{0,1\}^N$), and the matrix-vector multiplication can be implemented by the MAC operation of the proposed TD-CiM array as shown in Fig.\ref{fig:hdc}(a).
After the encoding, HDC superimposes together the encoded hypervectors corresponding to the same class of feature vectors to create a composite representation of a phenomenon of interest known as a ``model hypervector'', which is then classified as an entry of the learned model. 

During the inference, the HDC firstly encodes the input feature query to produce a query hypervector $\vec{H}$ by performing the MAC operation. 
Parallel similarity search is then performed over the model hypervectors through an associative search operation. Such associative search, i.e.,  the accumulation XNOR function between the query hypervector and the stored class/model hypervectors, can be implemented by the CAM operation of the proposed TD-CiM array storing the learned model as shown in Fig.\ref{fig:hdc}(a).
The similarity  ($\delta$)  between the query $\vec{H}$ and all class/model hypervectors is computed to find out the class with the highest similarity to the query hypervector. 
In binary representation, Hamming distance is measured as the similarity metric. 
W

\begin{table*}\scriptsize
    \centering
\begin{threeparttable}
    \setlength{\abovecaptionskip}{0.cm}
    \caption{Summary of  time-domain CiM approaches}
    \begin{tabular}{|>{\columncolor[rgb]{0.839,0.863,0.898}} c|c|c|c|c|c|>{\columncolor[rgb]{0.663,0.820,0.557}}c|}
        \hline
        \rowcolor[rgb]{0.839,0.863,0.898} ~ & Ref.~\cite{saito2021analog} & Ref.~\cite{luo2021energy} & Ref.~\cite{yang2021timaq} & Ref.~\cite{song2021tdsram} & Ref.~\cite{zhang2021time} & {\cellcolor[rgb]{0.663,0.820,0.557}}\textbf{This Work} \\ \hline
        Signal Domain & Voltage & Time & Time & Time & Time & \textbf{Time} \\ \hline
        Technology & 22nm & 14nm & 28nm & 40nm & 28nm & \textbf{40nm} \\ \hline
        Cell Size & 1T-1R & 2T-1FeFET & 20T+4MUX & 12T & 2T-1MRAM & \textbf{3T-2FeFET} \\ \hline
        Functions & MAC & MAC, Activation & MAC & MAC & \makecell{Mult, ADD,\\ Boolean Logic} & \textbf{MAC, Search} \\ \hline
        Application & \makecell{Image\\ recognition} & \makecell{Image recognition,\\Reinforcement\\ learning} & \makecell{Image\\ classification} & \makecell{Image\\ classification} & \makecell{Image\\ classification} & \textbf{\makecell{Face detection,\\Activity/Speech recognition}} \\ \hline
        Efficiency (TOPS/W) & 13700 & 51318 & 65.89 & 716 & 12.28 & \textbf{8563} \\ \hline
        Reconfigurable & No & No & No & No & No & \textbf{Yes} \\ \hline
    \end{tabular}
    \label{tab:benchmarking}
    \vspace{-2ex}
\end{threeparttable}
\end{table*}

Using the same FeFET TD-CiM, the same chip can conduct dynamic fine-grain resource allocation of binary MAC and CAM for different tasks. We have designed and used a cycle-accurate simulator based on PyTorch~\cite{paszke2019pytorch} which emulates the TD-CiM functionality during the HDC inference. 
Our tool receives the energy consumption and execution time of HDC key operations (i.e., binary MAC and CAM) using the proposed FeFET TD-CiM and then expands those values to compute the application-level energy consumption and time when searching each query. 
As shown in Fig.\ref{fig:hdc}(b), the breakdown of time and energy consumption when performing the speech, activity recognition and face detection demonstrates the varying demand of binary MAC and CAM. Fig.\ref{fig:hdc}(c) compares the energy consumption and execution time of the proposed TD-CiM for different applications compared to the GPU platform (GeForce RTX 3060 Ti GPU). It shows that our FeFET TD-CiM array in 32-stage configuration provides, on average 106$\times$ energy reduction and 63$\times$ speedup than GPU. Table \ref{tab:benchmarking} compares our proposed solution with other CiM solutions~\cite{saito2021analog,luo2021energy,yang2021timaq,song2021tdsram,zhang2021time}.
Due to the unique 2FeFET cell which can implement both the AND and XOR(XNOR) logic, our proposed FeFET TD-CiM allows fine-grain and highly flexible reconfigurability between the binary MAC and CAM functionalities using the same array. With the energy-efficiency of 8563 TOPS/W, the proposed FeFET based TD-CiM is a highly promising candidate processing fabric for various in-memory computing applications.


\vspace{-1ex}
\section{Conclusion}
\label{sec:conclusion}

In this work, we proposed a homogeneous processing fabric using FeFET based TD-CiM array that can support both the binary MAC and CAM. We have demonstrated the AND/XOR(XNOR) logic functionality of a 2FeFET cell and integrated such cells in a delay chain, which supports TD-CiM realization of binary MAC and CAM through a two-phase operation. 
Both theoretical and experimental validations are conducted and the robustness of the system against FeFET variation is demonstrated. 
We benchmark this FeFET TD-CiM processing fabric in the context of HDC, and  show that our system can adapt to different tasks with varying demands over the binary MAC and CAM through dynamic resource allocation. Our proposed FeFET
TD-CiM provides a promising CiM design with its versatility and high performance.

\vspace{-1ex}




\bibliographystyle{IEEE_tran}

\bibliography{bib}


\end{document}